\documentclass[superscriptaddress,showpacs,amssymb,10pt,reprint,aps,prd,longbibliography,nofootinbib,floatfix]{revtex4-1}

\usepackage{geometry}
\usepackage{graphicx,epsfig,amssymb}
\usepackage{amsmath,amsfonts, times}
\usepackage{bm}

\usepackage{graphicx}
\usepackage{subcaption}
\usepackage{float}

\usepackage{epstopdf}
\usepackage[linktocpage,colorlinks]{hyperref}
\usepackage[usenames]{color}

\usepackage{soul}
\usepackage[utf8x]{inputenc}
\usepackage[usenames]{color}
\usepackage{booktabs}
\usepackage{threeparttable}
\newcommand{\beq}{\begin{equation}}
\newcommand{\eeq}{\end{equation}}
\newcommand{\bqn}{\begin{eqnarray}}
\newcommand{\eqn}{\end{eqnarray}}
\begin{document}
	\title{Entanglement induced by quantum gravity in an infinite square well}
	\author{Chi Zhang}
	\email{zhangchi3244@gmail.com}
	\affiliation{School of information engineering, Zhejiang Ocean University, zhoushan, 316022, China}
	\author{Fu-Wen Shu}
	\email{shufuwen@ncu.edu.cn; Corresponding author}
	\affiliation{Center for Relativistic Astrophysics and High Energy Physics, Nanchang University, Nanchang, 330031, China}
\affiliation{GCAP-CASPER, Physics Department, Baylor University, Waco, Texas 76798-7316, USA}

\begin{abstract}
	In this work, we examine the entanglement dynamics of two massive particles confined within their respective infinite square potential wells induced by gravity. Assuming that each particle is initially in the ground state of its infinite  square well, we investigate the formation of entanglement between the particles as a result of  their gravitational interaction during their adiabatic movement towards each other. Our results reveal that, under suitable parameters, entanglement swiftly emerges. Compared to previous schemes, our approach significantly reduces the particle mass, greatly lowering the experimental threshold.

\end{abstract}
	\date{\today}
	\maketitle
\section{Introduction}
The debate over whether the gravitational field is inherently quantum in nature has persisted over time, primarily due to the absence of conclusive experimental evidence. Traditional wisdom posits that at macroscopic scales where gravity exerts significant influence, quantum mechanics might cease to apply \cite{Penrose:1996cv}, eliminating the need for conventional quantization of gravity. Recent emergent gravity models suggest an alternative perspective that questions the fundamental nature of gravity and its necessity for quantization  \cite{Jacobson:1995ab,Shu:2010nv}. However, classical gravitational concepts confront substantial hurdles, notably when addressing issues like black hole singularities and information loss paradox. Consequently, the current central inquiry revolves around the pursuit of experimental evidence to validate gravity's quantum nature.

Testing quantum nature of gravity, however, often necessitates stringent experimental protocols, as indicated by various studies \cite{r6}. Fortunately, a novel desktop experiment scheme called the quantum gravity induced entanglement of masses (QGEM)  has emerged in the low-energy realm, offering a more accessible approach \cite{Marletto:2017kzi,Bose:2017nin}. In this scheme, two massive particles initially placed in a superposition of two positions through a Stern-Gerlach experiment are brought together. If entanglement manifests between these particles, it signifies the quantum nature of the gravitational field, aligning with the Locally Operations and Classical Communication (LOCC) principle \cite{Guhne:2008qic}.

A key challenge in implementing the QGEM protocol is creating a large spatial superposition for a substantial mass object \cite{Bose:2017nin,vandeKamp:2020rqh,Toros:2020dbf}. Atom interferometers are usually adept at generating large spatial superpositions \cite{McGuirk:2002zz,Asenbaum:2016djh,Dimopoulos:2008hx}, but the mass is several orders less than what is required to test gravity's quantum nature. To date, the best scheme is the macromolecules, which can generate moderate-size superpositions  ($\sim 10$ nm to 1 µm) of masses around $10^{-19}–10^{-17}$ kg \cite{Romero-Isart:2011sdw,Sekatski_2014,Wan:2016vzv,Arndt2014TestingTL,Pino_2018}. However, this is still far from the requirement of the QGEM protocol. QGEM calls for a large spatial superposition $\sim O(10–100)$ µm of heavy masses ($m \sim O(10^{-17}–10^{-14}) $kg), with the Stern-Gerlach effect being the likely choice \cite{Keil:2020hvv,PhysRevLett.123.083601,Margalit:2020qcy,Marshman:2021wyk}.

In order for overcoming this difficulty, many schemes  have been proposed in the last few years. For example,  in \cite{r7}, the authors performed a scheme by placing two atomic gas interferometers next to each other in parallel and looking for correlations in the number of atoms at the output ports as a test of QGEM. Ref. \cite{Zhou:2022jug} presents a method of achieving a mass-independent enhancement of superposition via diamagnetic repulsion from current-carrying wires. Ref. \cite{Yant:2023smr} proposes a thought experiment where two particles are initially prepared in a superposition of coherent states within a common three-dimensional harmonic trap. And a scheme by using massless particles (a laser pulse) to generate the gravitational field \cite{He:2023hys} or by exploiting the two-phonon drive in a hybrid quantum setup \cite{Cui:2023ped}.

However, almost all the previous studies treated the particles in QGEM experiment more like tiny solid particles. What if they behave as a wave function and evolve according to Schrodinger's equation? Gravitational interactions in the Schrödinger equation may induce entanglement between different eigenstates.

In this work, we would like to introduce an alternative scheme. In this new scheme, we consider two adjacent one-dimensional infinite square potential wells and investigate the entanglement of two particles in each well under the influence of gravitational interaction between them. Specifically, we consider two heavy ultracold particles in the ground state in their respective potential wells. We investigate what the overall state before decoherence would be if we bring these two potential wells close to each other to a certain distance and gravity between them comes into play. 

\section{Setup}
\begin{figure}
\centering
\includegraphics[scale=0.3]{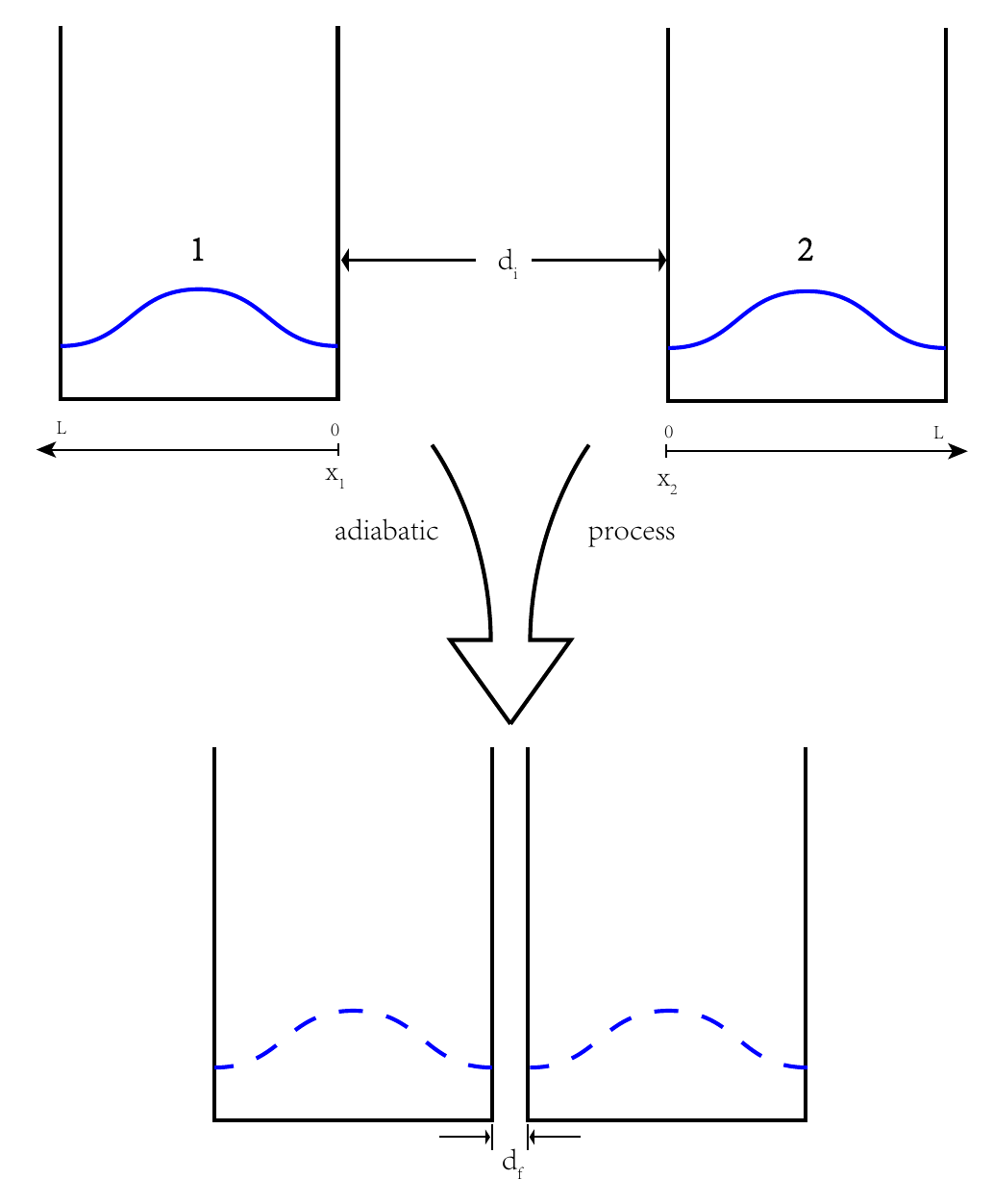}
\caption{Initially, two microscopic particles, each in its ground state, are separated by a distance $d_{ini}$. Subsequently, they undergo a gradual approach towards each other until they reach a final distance of $d_f$.  The entire process is executed at a sufficiently slow pace to ensure adiabaticity. The width of the square potential well is denoted by $L$.}\label{f1}
\end{figure}

In the initial step, we prepare two particles, each in the ground state of a one-dimensional infinite deep square potential well, as depicted in Fig.\ref{f1}. In order to avoid possible decoherence between the two particles, we initially prepare their ground state while keeping them at a significant distance from each other. 
The initial state of each particle can be expressed as:
\beq\label{1}
\left| {{\phi _{ini}}} \right\rangle  = \left| g \right\rangle,
\eeq
where $\left| g \right\rangle $ represents the ground state of a particle within the well and the subscript ``ini'' means ``initial''. The corresponding wave function and energy are given by:
\beq\label{2}
{\phi _{ini}}\left( x \right) = \left\langle {x}
\mathrel{\left | {\vphantom {x {{\phi _{ini}}}}}
\right. \kern-\nulldelimiterspace}
{{{\phi _{ini}}}} \right\rangle = \sqrt {\frac{2}{L}} \sin \left( {\frac{{\pi x}}{L}} \right),\ \ {E_{ini}} = \frac{{{\pi ^2}{\hbar ^2}}}{{2m{L^2}}}.
\eeq
And we label the i-th energy eigenstate of a single potential well as $\left| i \right\rangle$. Subsequently, we gradually bring the wells of the two particles closer until they are separated by a specific short distance. This process is performed slowly enough to ensure the applicability of the adiabatic approximation, thereby maintaining the particles in instantaneous energy eigenstates throughout. Precisely, we introduce a characteristic time scale $\tau_c$ to quantify the rate of change, defined as $\tau_c\equiv\frac{d_{ini}-d_f}{v}$, where $d_{ini}$ and $d_f$ represent the initial and final distances between the particles, respectively, and $v$ denotes their relative approaching velocity. The adiabatic approximation remains valid when $\tau_c$ is much larger than the characteristic time scale of oscillation, represented by $\tau_o\equiv \frac{2\pi\hbar}{E_{ini}}$.

The whole quantum system evolves under the Schr\"{o}dinger equation:
\beq\label{3}
i\hbar {\partial _t}\psi \left( {t,{\textbf{r}_1},{\textbf{r}_2}} \right) = \hat H\left( t \right)\psi \left( {t,{\textbf{r}_1},{\textbf{r}_2}} \right).
\eeq
The Hamiltonian could be divided into free Hamiltonian and interaction Hamiltonian:
\beq\label{4}
\hat H = {{\hat H}_0}\left( t \right) + {{\hat H}_{{\mathop{\rm int}} }}\left( t \right),
\eeq
where
\beq\label{5}
{{\hat H}_0}\left( t \right) =  - \sum\limits_{j = 1}^2 {\frac{{{\hbar ^2}}}{{2{m_j}}}\nabla _j^2}  + \sum\limits_{j = 1}^2 {{V_j}\left( t \right)}.
\eeq 
Although ${V_j}$ depends on $t$, due to the uniformity of the background spacetime, the adiabatic approach of the square potential wells will not change the quantum state. Instead, the time-dependent gravitational interaction during the approach process plays a decisive role in the instantaneous quantum eigenstate. In order to illustrate this more precisely, we establish the coordinate system shown in Fig.\ref{f1}.

In this coordinate system and assuming the two particles have the same mass, $m_1=m_2=m$:
\beq\label{16}
\begin{split}
&{V_j} = \left\{ \begin{array}{l}
	0,\;\;\;\;\;0 \le {x_j} \le L\\
	+ \infty ,\;\;{x_j} < 0,\;{x_j} > L
\end{array} \right.
\\
&\text{and}
\\
&{{\hat H}_{{\mathop{\rm int}} }}\left( t \right) =  - \frac{{G{m^2}}}{{\left| {{{\hat r}_1} - {{\hat r}_2}} \right|}} =  - \frac{{G{m^2}}}{{{x_1} + {x_2} + d\left( t \right)}},
\end{split}
\eeq
where $d\left( t \right)$ represents the instantaneous distance of the potential wells during the adiabatic approach process.

Note that ${{\hat H}_{{\mathop{\rm int}} }}\left( t \right)$ is time-dependent due to the gradual approach of the two wells towards each other and it includes position operator ${\hat r_{1/2}}$. It is the zeroth-order term of quantum gravitational interaction, and its derivation can be found in \cite{Bose:2022uxe,Marshman:2019sne}. It differs from the usual Schr\"{o}dinger-Newton Hamiltonian where the gravitational field is classical and cannot  induce entanglement \cite{Ligez:2021fnu}:
\beq\label{6}
{{\hat H}_{SN}}^\prime  =  - G{m^2}\sum\limits_{i,j = 1}^2 {\left\{ {\int {\prod\limits_{k = 1}^2 {{d^3}{r_k}^\prime } } } \right\}\frac{{{{\left| {\psi \left( {t,{r_1}^\prime ,{r_2}^\prime } \right)} \right|}^2}}}{{\left| {{r_i} - {r_j}^\prime } \right|}}} .
\eeq

In our numerical analysis, we begin by considering that, since the particles initially separated by a significant distance and gravitational interaction does not play a role, the initial common ground state of the two particles is a direct product state:
\beq\label{7}
\left| {{\psi _{ini}}} \right\rangle = \left| {{g_1}} \right\rangle \otimes \left| {{g_2}} \right\rangle.
\eeq
Subsequently, we adiabatically bring the particles closer together. At each instantaneous moment, each instantaneous eigenstate satisfies the time-independent Schr\"{o}dinger equation:
\begin{widetext}
\beq\label{9}
\begin{split}
	& - \frac{{{\hbar ^2}}}{{2m}}{\nabla _1}^2\phi \left( {{x_1},{x_2}} \right) - \frac{{{\hbar ^2}}}{{2m}}{\nabla _2}^2\phi \left( {{x_1},{x_2}} \right) - \frac{{G{m^2}}}{{{x_1} + {x_2} + d{\left( t \right)}}}\phi \left( {{x_1},{x_2}} \right) = E\phi \left( {{x_1},{x_2}} \right),\\
	&\phi \left( {{x_1},{x_2}} \right) = 0,\;\;\;{\rm{when}}\;\;{x_1} = 0,\;L\;{\rm{or}}\;{x_2} = 0,\;L,
\end{split}
\eeq
\end{widetext}
where we have used the coordinate system established as shown in Fig.\ref{f1}. For simplicity, in what follows we consider two particles of equal mass, specifically $1 \times {10^{-17}}$ kg, within a potential well width of 50 $\mu$m for the following numeric simulations. Without loss of generality, we focus on the eigenstates of the total Hamiltonian \eqref{4} associated with the lowest six energy levels, $n_1,n_2,\cdots n_6$. 

\begin{figure}\centering
	\includegraphics[scale=0.4]{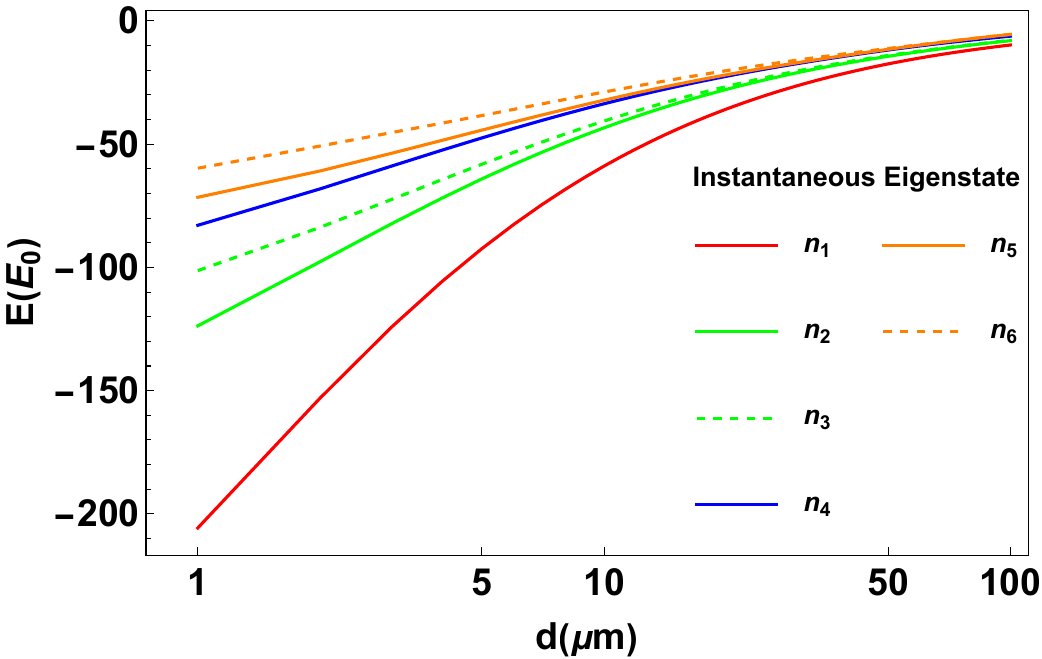}
	\caption{Adiabatic evolution curve of two microscopic particles of six different initial states. The horizontal axis represents the instantaneous distance between potential wells, while the vertical axis represents the instantaneous eigenvalue of energy for the particle system, measured in units of the ground state energy ${E_0} = {\pi ^2}{\hbar ^2}/m{L^2}$.  $n_1,n_2,\cdots n_6$ denote the lowest six energy levels whose corresponding eigenstates are $
{\left| {n_1} \right\rangle _{ini}} = \left| 1 \right\rangle \left| 1 \right\rangle,\ {\left| {n_2} \right\rangle _{ini}} = \frac{1}{\sqrt{2}}\left( {\left| 1 \right\rangle \left| 2 \right\rangle  + \left| 2 \right\rangle \left| 1 \right\rangle } \right),
 {\left| {n_3} \right\rangle _{ini}} = \frac{1}{\sqrt{2}}\left( {\left| 1 \right\rangle \left| 2 \right\rangle  - \left| 2 \right\rangle \left| 1 \right\rangle } \right),\ {\left| {n_4} \right\rangle _{ini}} = \left| 2 \right\rangle \left| 2 \right\rangle,
  {\left| {n_5} \right\rangle _{ini}} =  \frac{1}{\sqrt{2}}\left( {\left| 1 \right\rangle \left| 3 \right\rangle  + \left| 3 \right\rangle \left| 1 \right\rangle } \right),\
   {\left| {n_6} \right\rangle _{ini}} = \frac{1}{\sqrt{2}}\left( {\left| 1 \right\rangle \left| 3 \right\rangle  - \left| 3 \right\rangle \left| 1 \right\rangle } \right).$
}\label{eigencurve}
\end{figure}

Fig. \ref{eigencurve} shows that the adiabatic evolution curves of two microscopic particles for different initial states. The gravitational interaction between the particles becomes most pronounced as the potential wells approach each other closely. Additionally, as observed in the figure, all six eigenstates are associated with negative energy values. This is a consequence of the Newtonian potential being negative. Furthermore, as the separation distance ($d$) decreases, the evolution curve descends continuously but never intersects, which is consistent with the fact that the gravitational potential increases as $d$ decreases.

Due to the particle exchange invariance of the Hamiltonian in Eq. \eqref{9}, the eigenstates exhibit either symmetric or anti-symmetric behavior under particle exchange. This characteristic is also evident from Fig. \ref{eigencurve}. For curves $n_1$ and $n_4$, where the initial state ($| n_1\rangle_{ini}$ and $| n_4\rangle_{ini}$ ) is a direct product of two identical states, no splitting occurs as $d$ decreases. However, for $n_2$ and $n_3$, composed of two distinct states, the initial states ($| n_2\rangle_{ini}$ and $| n_3\rangle_{ini}$ ) are either symmetric or anti-symmetric. Initially, these states have degenerate energies in the absence of gravitational effects, but as gravitational influence becomes significant, their energy levels split. Consequently, we observe the dashed green line (representing the anti-symmetric state) and the solid green line (representing the symmetric state) undergo a split as $d$ decreases, with the anti-symmetric state having a higher energy than the symmetric state. This pattern is also observed for $n_5$ and $n_6$.

\begin{figure}\centering
	\includegraphics[scale=0.4]{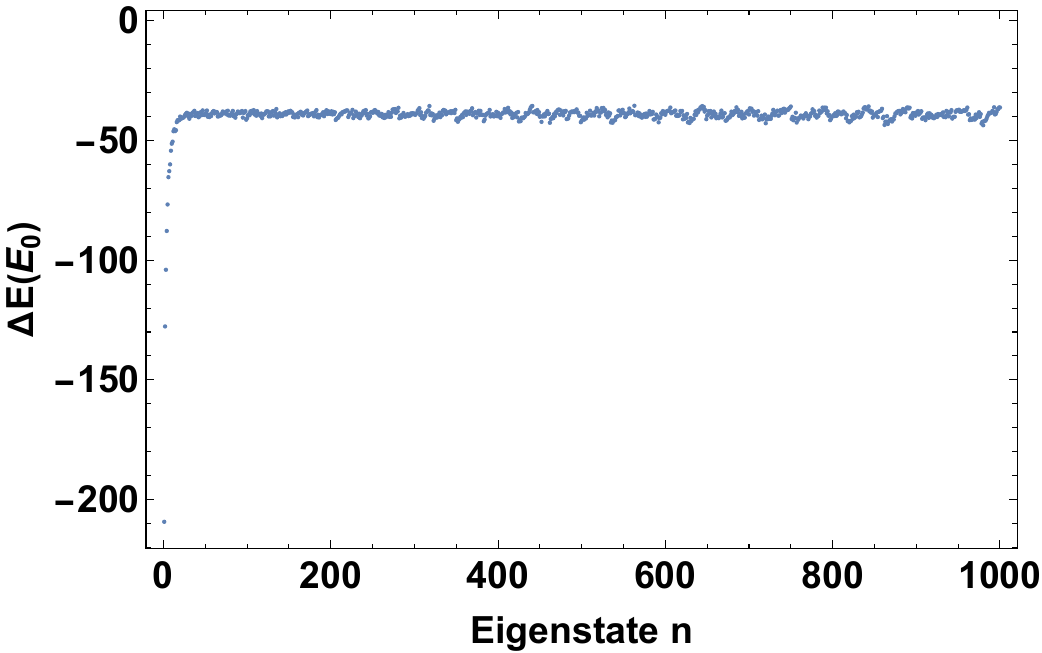}
	\caption{Energy eigenvalue changes of the lowest 1000 energy levels of the final state, where $m_1=m_2=m=10^{-17}\text{kg}$, $L=50 \mu$m, and $d_f=1\mu$m. The energy change has been measured in units of the ground state energy ${E_0} = {\pi ^2}{\hbar ^2}/m{L^2}$.}\label{f6}
\end{figure}
The final state of the adiabatic evolution, marking the point where the two potential wells cease approaching each other, is depicted in Fig. \ref{f6}. In this state, the separation distance between the wells, denoted as $d_f$, is set to $1\mu$m. The horizontal coordinate $n$ represents the 1000 lowest energy eigenstates, starting from the ground state, while the vertical coordinate is calculated as:

\beq\label{8}
\Delta E = {E_{f}} - {E_{ini}} = {E_{f}} - \frac{{{n_1}^2{\pi ^2}{\hbar ^2}}}{{2m{L^2}}} - \frac{{{n_2}^2{\pi ^2}{\hbar ^2}}}{{2m{L^2}}},
\eeq
where $E_i$ and $E_f$ denote, respectively, the energy for initial state and final state of the two-particle system.

The ground state exhibits the most significant change in energy. As the energy level, denoted by $n$, increases, the energy shift decreases rapidly and tends towards a stable value. As expected, this energy shift, denoted as $\Delta E$, approaches the asymptotic value of $\Delta E \sim - \frac{{G{m^2}}}{L}$, which reflects the fact that higher energy eigenstates dynamics are primarily determined by the confining potential, with the gravitational interaction having a reduced impact on the dynamics and so the only energy change is due to the classical gravitational interaction.

\section{Entanglement detection}
It is of great significance to investigate to what extent the entanglement can be formed at each energy eigenstate during the two particles' adiabatic evolution towards the final state. With the combination of the eigenstates of their respective confining infinite square well, the entangled final state could be described as:
\beq\label{11}
\left| {{\psi _f}} \right\rangle  = {a_{11}}\left| {11} \right\rangle  + {a_{12}}\left| {12} \right\rangle  +  \cdots {a_{ij}}\left| {ij} \right\rangle  \cdots  + {a_{nn}}\left| {nn} \right\rangle ,
\eeq
where $n$ is the highest excitation energy level we assume. In the appendix we show that $n=100$ is accurate enough for our numerical calculations. As a result, we are safe to set $n=100$ for our numerical simulation. In fact, $n$ is just an artificially selected simulation calculation parameter. When many high-energy states are not excited as shown in Appendix A (Fig. \eqref{p1}), the entangled quantum state can be regarded as a quantum state with lower dimension.

\begin{figure}\centering
	\includegraphics[scale=0.4]{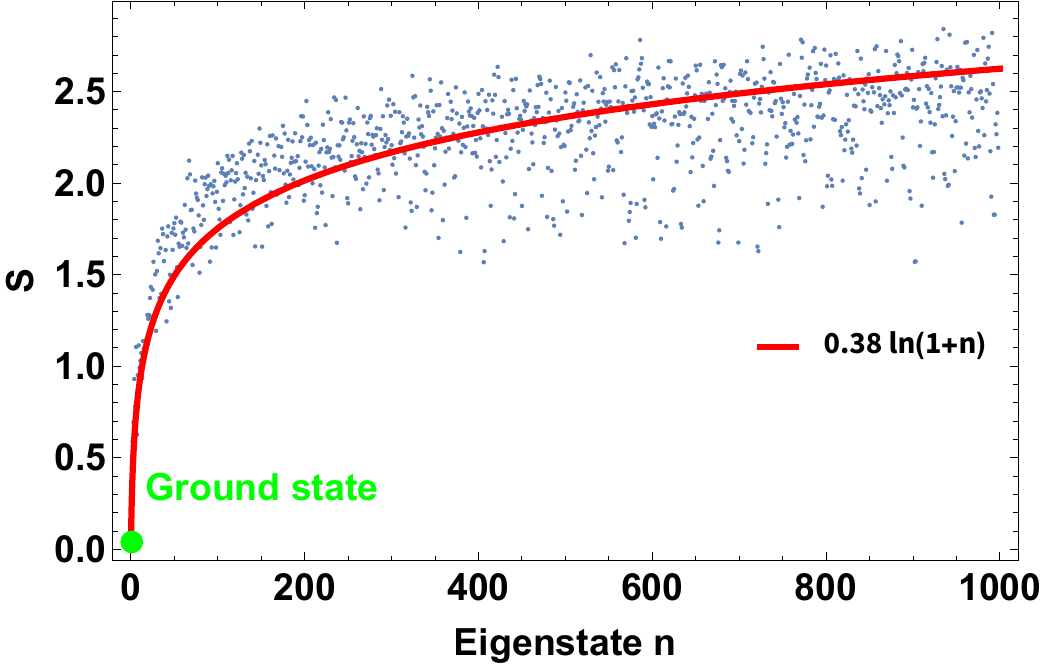}
	\caption{Entanglement entropy of the lowest 1000 energy levels of Hamiltonian \eqref{4} in the final state, where $m_1=m_2=m=10^{-17}\text{kg}$, $L=50 \mu$m, and $d_f=1\mu$m.}\label{f7}
\end{figure}

To show more details of the entanglement formation, it is helpful to calculate the entanglement entropy, which is defined as
\beq\label{10}
S({\rho _A}) = S({\rho _B}) =  - {\rm{tr}}\left( {{\rho _A}\ln {\rho _A}} \right) =  - {\rm{tr}}\left( {{\rho _B}\ln {\rho _B}} \right),
\eeq
where ${\rho _A}$ and ${\rho _B}$ are the reduced density matrices of particles A and B, respectively.  Remarkably, the changes in entanglement entropy for each energy eigenstate of Hamiltonian \eqref{4} exhibit completely contrasting behaviors. In general, as the energy level of the eigenstate increases, the degree of entanglement between the two particles also increases. Fig. \ref{f7} demonstrates this trend, where the entanglement entropy ($S$) initially experiences rapid growth before reaching a slower rate of increase. Fitting these scattered data points reveals the characteristic pattern of a logarithmic function. This behavior arises from the fact that in a one-dimensional infinite deep square potential well, the energy level of an independent particle is proportional to the energy level number ($n$). As $n$ increases, both the energy values and the intervals between adjacent energy levels expand. According to the first-order perturbation formula of quantum mechanics, a larger number of distinct eigenstates of confining well participate in the first-order perturbation correction of the eigenstate, leading to an increase in entanglement. Additionally, unlike the energy correction $\Delta E$ depicted in Fig. \ref{f6}, the entanglement entropy exhibits greater fluctuations and a scattered distribution. In what follows, we will primarily focus on the entanglement formation of the ground state, given its ease of experimental manipulation and inherent stability. From Fig. \ref{eigencurve}, we can see that the quantum state curve, ${n_1}$, whose initial conition is the ground state Eq. \eqref{7}, will remain in the ground state throughout the adiabatic evolution until the end.

\begin{figure}
	\centering
	{\includegraphics[scale=0.4]{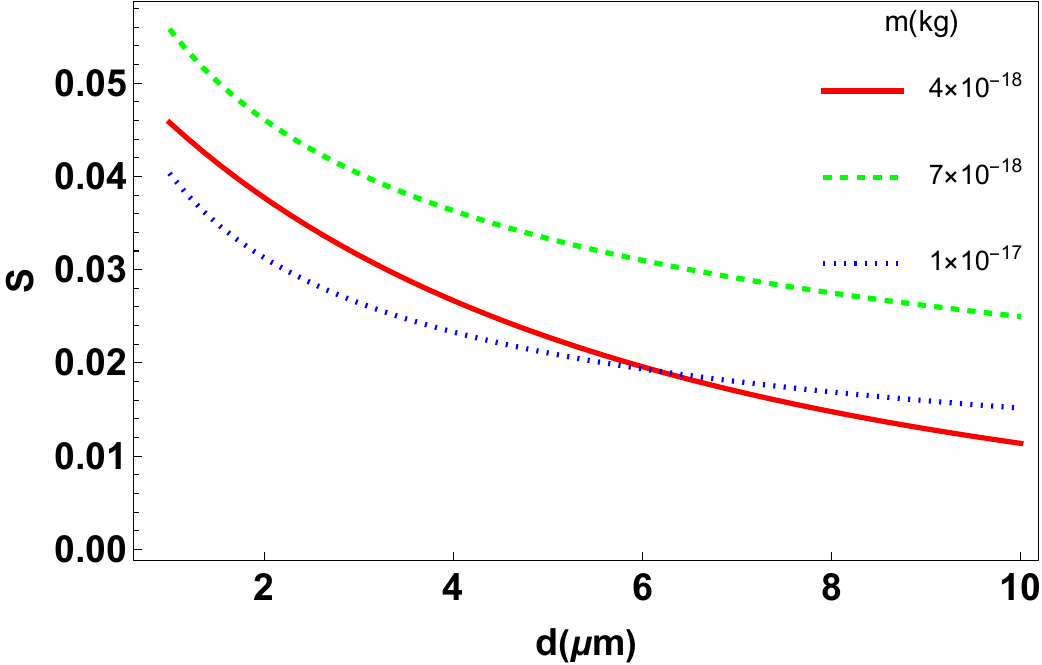}}
	\caption{When $L$ is fixed at $50\mu $m, the entanglement entropy of the ground state as a function of the separation distance between potential wells during adiabatic approach.}\label{dsm}
\end{figure}

\begin{figure}
	\centering
	\subcaptionbox{Entanglement entropy varies with $m$ and $L$.\label{f11a}}
	{\includegraphics[scale=0.25]{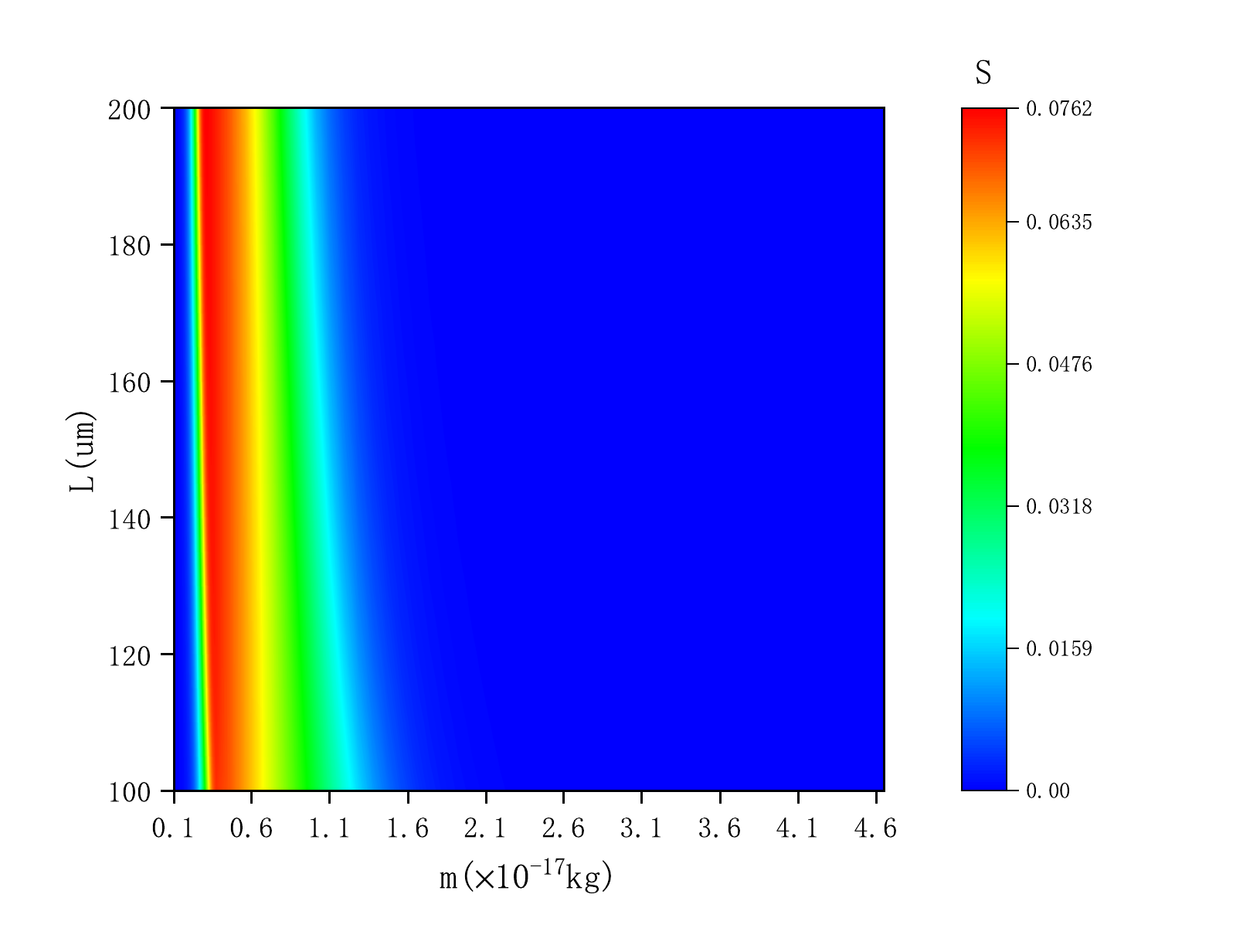}}
	\subcaptionbox{Entanglement witness varies with $m$ and $L$.\label{f11b}}
	{\includegraphics[scale=0.25]{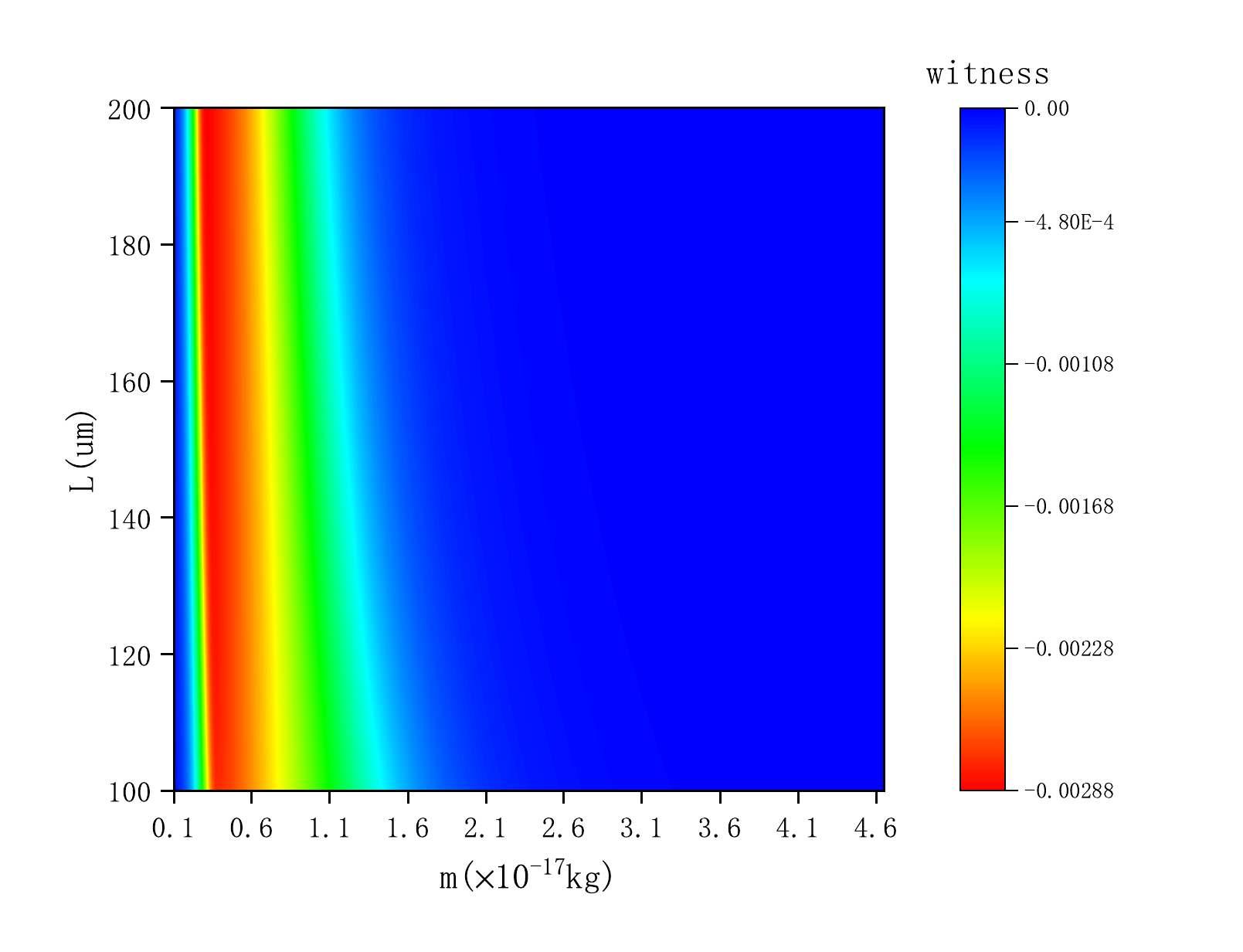}}
	\caption{Contour plots of entanglement entropy and entanglement witness with respect to $m$ and $L$ for the final state. In both subfigures, $d$ is fixed at $1\mu$m in the final state.}\label{f11}
\end{figure}

Next, let us explore how entanglement entropy is influenced by system parameters, such as potential well width, particle mass, and the distance between two potential wells. Intuitively, these factors play a crucial role in shaping the entangled state. Given the gravitational effect's prominence during the adiabatic evolution of two potential wells into their final state, we expect the induced entanglement to be maximized in the final state. This observation is corroborated by Fig. \ref{dsm}. Fig. \ref{dsm} shows that as the potential well spacing $d$ decreases, the gravitational interaction becomes more significant. The entanglement entropy of the ground state with an initial entanglement entropy 0 increases as $d$ decreases, thus verifying quantum gravity inducing entanglement formation. Hence, all subsequent calculations assume the wells are in this final adiabatic state. The numerical results on the effects of particle mass and potential well width on the degree of entanglement are presented as counter plots in Fig. \ref{f11}, where brighter colors indicate a higher level of entanglement, and vice versa.

Clearly, in Fig. \eqref{f11a}, the parameter $m$ exerts a significant influence on the entanglement, with the entanglement entropy reaching its maximum at a critical mass of $m \approx 4.8 \times 10^{-18}$ kg. When $m$ exceeds this critical value, $S$ exhibits a slight decrease as $L$ increases. Conversely, for $m$ below the critical mass, $S$ experiences a slight increase as $L$ increases. A stronger gravitational interaction induced by a particle mass exceeding the critical mass does not necessarily lead to greater quantum entanglement, which appears highly counterintuitive. This phenomenon is related to the modification of the two-particle eigenstates under gravitational interaction by the confining potential wells. Similar to the wavefunction of a hydrogen atom, the two-particle system under gravitational interaction possesses an eigenstate. When the gravitational interaction becomes sufficiently strong, the spatial extent of the wavefunction decreases, leading to a significantly reduced influence of the potential wells on the wavefunction. Consequently, the quantum entanglement between the two particles diminishes. Notably, the critical mass is significantly smaller than the mesoscopic particle mass requirement of $1 \times 10^{-14}$ kg mentioned in reference \cite{Bose:2017nin}.

To experimentally detect entanglement, an entanglement witness is required. In the case of $n\otimes n$ systems, there exists a notable fidelity entanglement witness denoted as \textit{w}. This witness can be defined as follows \cite{Horodecki:2009zz}:
\beq\label{12}
\textit{w}(P_n^ + ) = {n^{ - 1}}I - P_n^ +  ,
\eeq
where $P_n^ +  = |\Phi _n^ + \rangle \langle \Phi _n^ + |$, $\left| {\Phi _n^ + } \right\rangle  = \frac{1}{{\sqrt n }}\sum\limits_{i = 1}^{n} {|i\rangle  \otimes |i\rangle }$ and $n = \dim {{\cal H}_A}$.
If $\left\langle {{\psi _f}} \right|\textit{w}\left| {{\psi _f}} \right\rangle < 0$, we can conclude that the state $\left| {{\psi _f}} \right\rangle $ is an entangled state. For the experimental implementation of the witness, it is necessary to decompose it into a number of local von Neumann (or projective) measurements \cite{Bourennane:2004uiy,r27}.

In the final ground state, all values of \textit{w} selected in Fig. \eqref{f11b} are negative. Notably, \textit{w} exhibits a similar changing pattern as $S$, indicating that regions with higher entanglement correspond to more negative values of \textit{w}, and vice versa. As indicated in \cite{Sun:2023ptw}, some renormalized entanglement witness, such as \textit{w} here, can not only detect  entanglement but also quantify the entanglement. From the perspective of entanglement detection, our preferred parameter space lies within the bright (red) strip marking the critical region in Fig. \eqref{f11b}, which encompasses smaller masses and larger potential well widths.

\begin{figure}
	\centering
	\subcaptionbox{The entangled final state with initial state ${\left| {n_1} \right\rangle _{i}}$.\label{f12a}}
	{\includegraphics[scale=0.4]{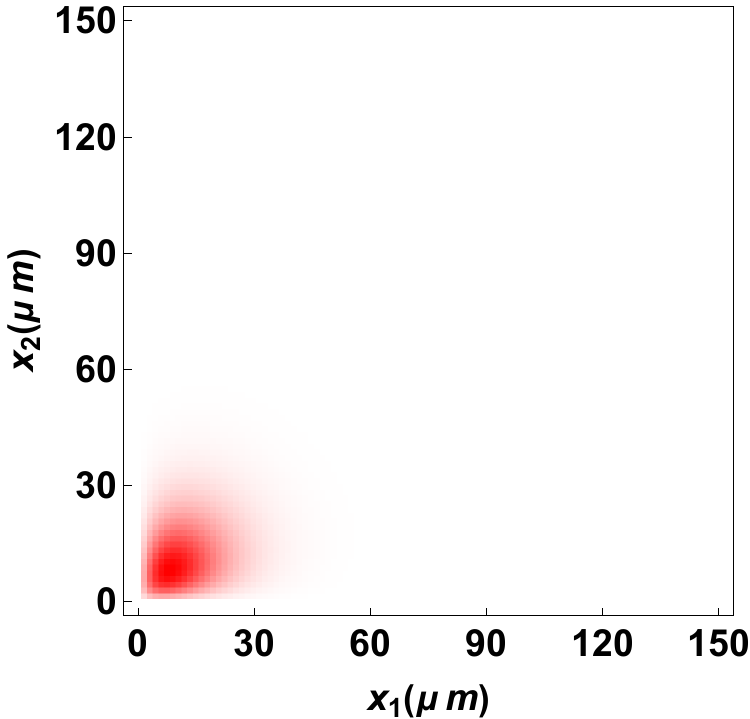}}
	\subcaptionbox{The entangled final state with initial state  ${\left| {n_3} \right\rangle _{i}}$.\label{f12b}}
	{\includegraphics[scale=0.4]{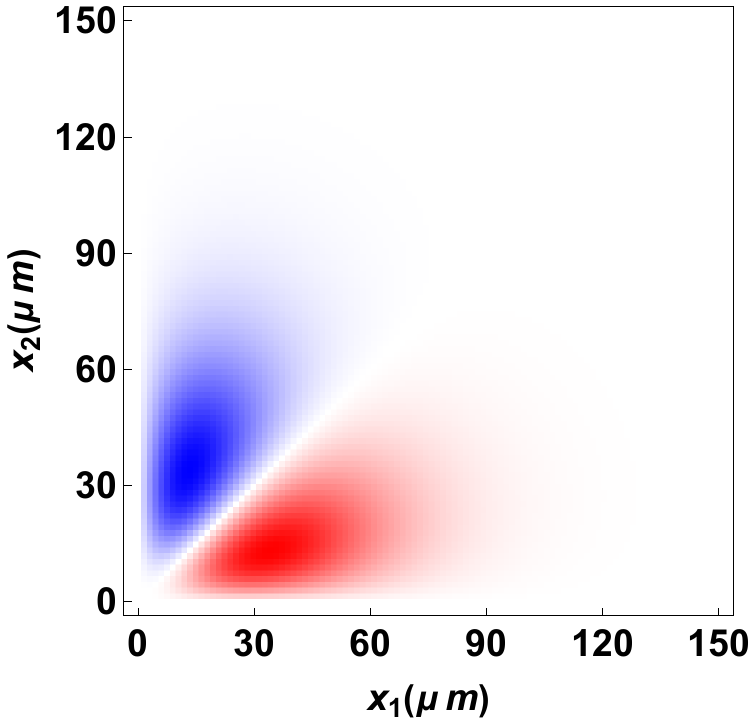}}
	\caption{The entangled final state in the position basis. White corresponds to zero wavefunction, red is positive ${\psi _f}\left( {{x_1},{x_2}} \right) > 0$, and blue is negative ${\psi _f}\left( {{x_1},{x_2}} \right) < 0$. In both subfigures, $d$ is fixed at $1\mu$m, $m$ is fixed at $4.8 \times 10^{-18} $kg and  $L$ is fixed at $150\mu$m.}\label{f12}
\end{figure}

All the previous analyses were carried out using the energy eigenstates of the Hamiltonian \eqref{5} corresponding to the confining potential well as the bases. However, entanglement does not depend on the choice of basis. To illustrate the entangled state in the position basis, we have plotted Fig. \ref{f12}. 

As expected, the symmetric and anti-symmetric states in position space correspond one-to-one with the particle exchange symmetry of the initial state. Due to the presence of quantum entanglement, the two-particle position correlation function exhibits quantum correlations, $\left\langle {{{\hat x}_1}{{\hat x}_2}} \right\rangle  \ne \left\langle {{{\hat x}_1}} \right\rangle \left\langle {{{\hat x}_2}} \right\rangle $.
	
However, it is worth noting that the aforementioned parameter selection just represents a possible theoretical choice. In our previous analysis, we have only considered the gravitational interaction between the two neutral particles. But in reality, other forces such as electric dipole interactions, including van der Waals and Casimir forces \cite{Klimchitskaya:2015vra, r19, Emig:2001dx}, also come into play. In Refs. \cite{Bose:2017nin, Marletto:2017kzi}, it was observed that shielding these interference forces requires maintaining a minimum distance between the nearest particles. Therefore, in real experiments, in addition to preparing a ground state that satisfies the necessary conditions, it is crucial to shield the electric dipole interactions between the particles to ensure that entanglement is solely caused by their gravitational fields.
 
Additional techniques, such as leveraging Feshbach resonances \cite{r24, r25} or implementing a conducting membrane \cite{vandeKamp:2020rqh, Westphal:2020okx}, have the possibility to further alleviate electromagnetic interactions between neutral atoms or particles. These approaches provide a broader range of parameter choices, facilitating the design and implementation of the experiment.

Moreover, according to the ultracold atom theory, a power law interaction potential between quantum systems can be treated as a perturbation pseudopotential resembling a delta function when the following two conditions are satisfied \cite{Busch:1998cey, r5, r21}.
\beq\label{13}
\begin{split}
	{\rho ^{ - 1/3}} \gg b,\\
	kb \ll 1.
\end{split}
\eeq
In Eq.\eqref{13}, the variable $\rho$ denotes the average particle density within the potential well. The parameter $b$ corresponds to the finite range of interaction forces between particles, and $k$ represents the average wave number of the particles
\beq\label{15}
k = \frac{{\left\langle p \right\rangle }}{\hbar } = \frac{{\sqrt {2m\left\langle {{E_{kin}}} \right\rangle } }}{\hbar }.
\eeq
If these two conditions are satisfied, the electric dipole interaction Hamiltonian can be expressed as follows:
\beq\label{14}
{{\hat H}_{dip}} \propto C\;\delta \left( {{x_1} - {x_2}} \right).
\eeq

${{\hat H}_{dip}}$ works only when the two wave functions overlap in space. However, the wave functions of the two particles are currently confined within their individual potential wells, with a separation of several microns between the wells. As a result, the electric dipole interaction between these neutral atoms has minimal influence on the entanglement induced by gravity. To ensure the fulfillment of these two conditions, we should carefully select suitable ranges for the parameters $m$ and $L$. In general, $\rho, k$ don't depend on $m$ and $\rho  \propto {L^{ - 1}}, k \propto {L^{ - 1}}$. So $L$ tends to take large values.

The quantum entanglement survives also for states above the ground state from Fig.\eqref{f7}, and hence particularly for thermal states, which means cooling to the ground state is not required as such.

\section{Decoherence in potential wells}
For our purpose, the experimental device should be in a weightless, vacuum, low temperature, and isolated environment. But due to the inevitable emission of thermal radiation and environmental scattering, the quantum system will undergo decoherence. Here we focus on the decoherence caused by environmental background radiation and make an upper bound estimate of the decoherence time. To estimate the expected decoherence we model the particles' environment by an (Ohmic) bath of harmonic oscillators. Then the reduced density matrix of the present system evolves according to the Caldeira–Leggett Master Equation \cite{clme,clme1}:
\begin{widetext}
	\beq\label{dec1}
	\begin{split}
			\frac{{\partial {\rho _S}({x_1},{x_2},{x_1}^\prime ,{x_2}^\prime ,t)}}{{\partial t}} &= \frac{{ - i}}{\hbar }\left[ {\hat H,{\rho _S}(t)} \right] + \left( {\frac{{i{m_1}{\gamma _0}\Lambda }}{\hbar }\left( {{x_1}^2 - {x_1}{{^\prime }^2}} \right) + \frac{{i{m_2}{\gamma _0}\Lambda }}{\hbar }\left( {{x_2}^2 - {x_2}{{^\prime }^2}} \right)} \right.
		\\
		& - \left. {\frac{{2{m_1}{\gamma _0}{k_B}T}}{{{\hbar ^2}}}{{\left( {{x_1} - {x_1}^\prime } \right)}^2} - \frac{{2{m_2}{\gamma _0}{k_B}T}}{{{\hbar ^2}}}{{\left( {{x_2} - {x_2}^\prime } \right)}^2}} \right){\rho _S}({x_1},{x_2},{x_1}^\prime ,{x_2}^\prime ,t),
	\end{split}
	\eeq
\end{widetext}	
Here ${\gamma _0}$ is the frequency-independent phenomenological damping constant, $\Lambda$ is the high-frequency cutoff constant and T is the temperature of environment. The term $\;\left( { - \frac{{2{m_1}{\gamma _0}{k_B}T}}{{{\hbar ^2}}}{{\left( {{x_1} - {x_1}^\prime } \right)}^2} - \frac{{2{m_2}{\gamma _0}{k_B}T}}{{{\hbar ^2}}}{{\left( {{x_2} - {x_2}^\prime } \right)}^2}} \right){\rho _S}$ describing the particles' spatial localization and the decay of off-diagonal elements in position basis corresponds to a decoherence rate. This decoherence model is valid when the environment space is large and the system-environment coupling is extremely weak and the memory effect of the environment decays quickly enough.

Additionally, air molecules remaining in vacuum, particle thermal radiation and particle intrinsic degrees of freedom are also sources of decoherence and will cause decoherence more quickly. As decoherence  governed by Eq. \eqref{dec1} proceeds, the purity of the particle system, $\text{tr}\left( {{\rho_S ^2}} \right)$, exhibits an exponential decay  over time. The decoherence characteristic time, denoted as  $\tau_d$, can be defined as $\frac{d}{dt}\operatorname{tr}(\rho^2)\sim-\frac{1}{\tau_d}\operatorname{tr}(\rho^2)$. When $\tau_d$ is longer than the time required for experimental preparation and detection ($\sim \tau_c$), we can observe the anticipated phenomenon of gravity-induced quantum entanglement. In order to estimate $\tau_c$, a more precise calculation of the adiabatic process is needed in the future. This will allow us to optimize the experimental parameters and procedures to ensure that the required experimental time satisfies this relation.

\section{Potential experimental Proposal}
\begin{figure}\centering
	\includegraphics[scale=0.32]{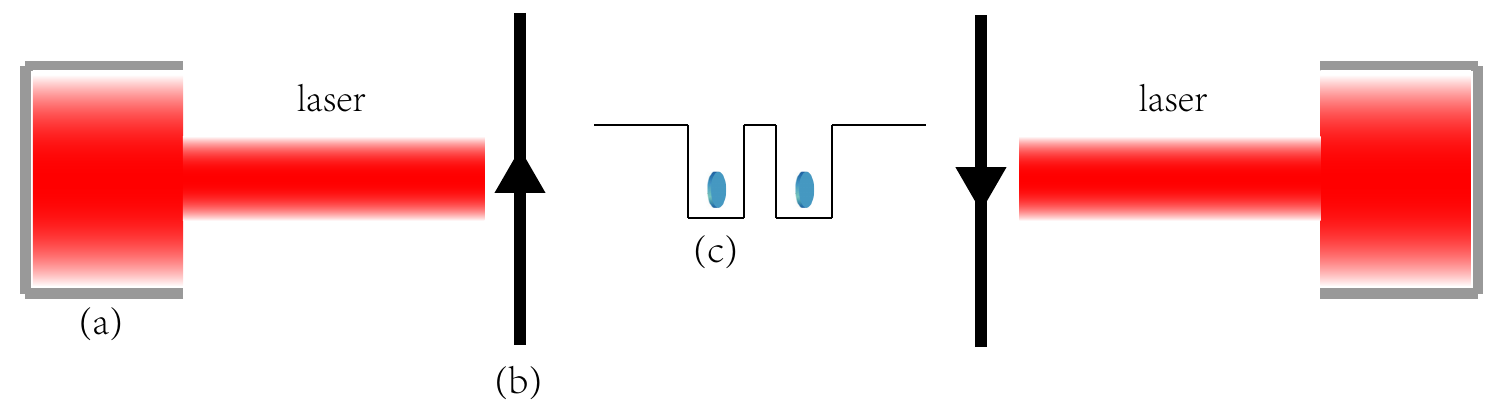}
	\caption{Schematic of proposed experiment showing (a) optical access, (b) anti-Helmholtz configuration, (c)levitated superconducting osmium particles.}\label{set}
\end{figure}

The proposed entangled state can be realized by confining two micron-sized superconducting, magnetically levitated osmium mirrors within two optical potential wells, as illustrated schematically in Fig. \ref{set}.

The metal osmium particles, which have a superconducting critical temperature of ${T_{\rm{c}}} = 700\text{mk}$, repel the magnetic field generated by the anti-Helmholtz coil and float in mid-air due to the Meissner effect. This arrangement minimizes photon scattering and absorption within the optically levitated system \cite{Grossardt:2015moa}, enabling the quantum state to be highly isolated from the environment and achieve long coherence times.

Osmium is a dense material, with a density of  $\rho  = 22.57\text{g}\cdot\text{cm}^{-3}$, allowing for relatively small particle volumes with significant mass. Furthermore, the osmium particles are shaped into disks to reduce their horizontal dimensions. Another potential decoherence pathway involves the absorption of blackbody radiation. However, for radiation frequencies below the superconductor's gap energy, absorption is negligible. Thus, by maintaining an environmental temperature significantly lower than the superconductor's critical temperature, the osmium particles absorb very few ambient photons, thereby mitigating decoherence through this mechanism \cite{r26}.

To simulate an infinitely deep square potential well, we utilize an optical potential well created by a laser standing wave. By continuously modulating the laser signal, the two potential wells adiabatically approach each other and evolve into the final state.

\section{conclusion}
In this paper, we propose a new approach to probing QGEM through a two-potential-well quantum system. By harnessing the gravitational fields of particles trapped within these wells, we achieve an entangled final state, vividly demonstrating the quantum essence of gravity.  Compared to the original scheme proposed in \cite{Bose:2017nin, Marletto:2017kzi}, our proposal offers notable advantages. It eliminates the need for spin particles, Stern-Gerlach devices, or the preparation of superposition states in space, simplifying the experimental procedure. It is also expected to reduce the impact of electromagnetic interference. Additionally, the required mass for achieving maximum entanglement in the ground state, $4.8 \times 10^{-18}$ kg, is significantly smaller than the mass requirement stated in the previous studies. Therefore, it prolongs the decoherence time \cite{r28}, prolongs the gravitationally induced collapse time \cite{r29}, and lowers the experimental threshold.  Moreover, as long as the coherence of the final state is maintained, entanglement persists without the need for additional time to form it.

Our numerical simulations reveal three pivotal factors that govern the entanglement of the final state:  particle mass ($m$), potential well width ($L$), and potential well spacing ($d$). Among them, the particle mass has the most significant impact. It induces maximum entanglement at a certain critical value. Conversely, as the potential well width and spacing increase, the entanglement entropy ($S$) generally decreases, albeit with slight elevations observed at specific intervals.

The energy level distribution within a potential well intricately mirrors its unique attributes. Exploring alternative potential wells, like spherical or harmonic ones, can significantly reshape the resultant entangled state and its parameter sensitivity, potentially unlocking new avenues for discovering more entangled eigenstates and lowering the necessary particle mass threshold. This warrants further investigation.

\section*{Acknowledgments}
We would also like to extend our thanks to  Anupam Mazumdar for carefully reading the manuscript and for providing valuable suggestions and comments during the course of this work. This work was supported by the National
Natural Science Foundation of China under the Grant No. 12375049, and Key Program of the Natural
Science Foundation of Jiangxi Province under the Grant No. 20232ACB201008.

  \begin{appendix}
  \section{Energy level with different n}\label{app-a}
\begin{figure}[h]\centering
	\includegraphics[scale=0.4]{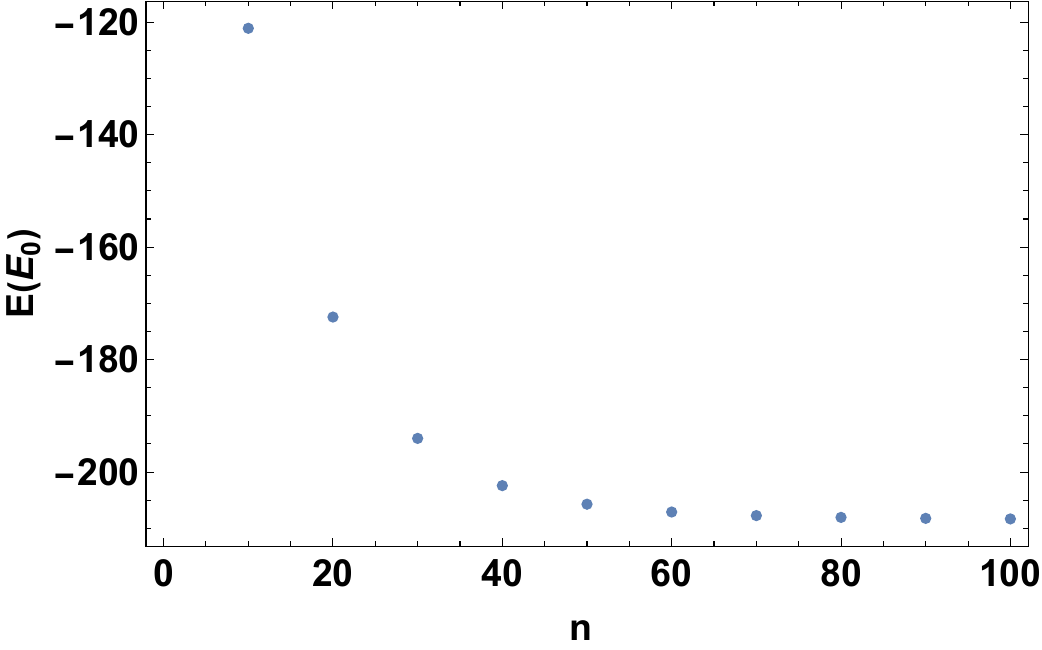}
	\caption{Energy eigenvalue of the ground state foe different n(nmax), where $m_1=m_2=10^{-17}\text{kg}$, $L=50 \mu$m, and $d_f=1\mu$m. The energy change has been measured in units of the ground state energy ${E_0} = {\pi ^2}{\hbar ^2}/m{L^2}$}.\label{nmax}
\end{figure}

We utilize Fourier spectrum method with several different n to calculate the energy eigenvalue of the ground state of intrinsic Eq. \eqref{9}.
In Fig. \ref{nmax} the calculated energy is already fairly stable when n reaches 100, which demonstrates the simulation has been quite accurate.

\begin{figure}
	\centering
	\subcaptionbox{$m = 1 \times {10^{ - 17}}kg$\label{p1}}
	{\includegraphics[scale=0.4]{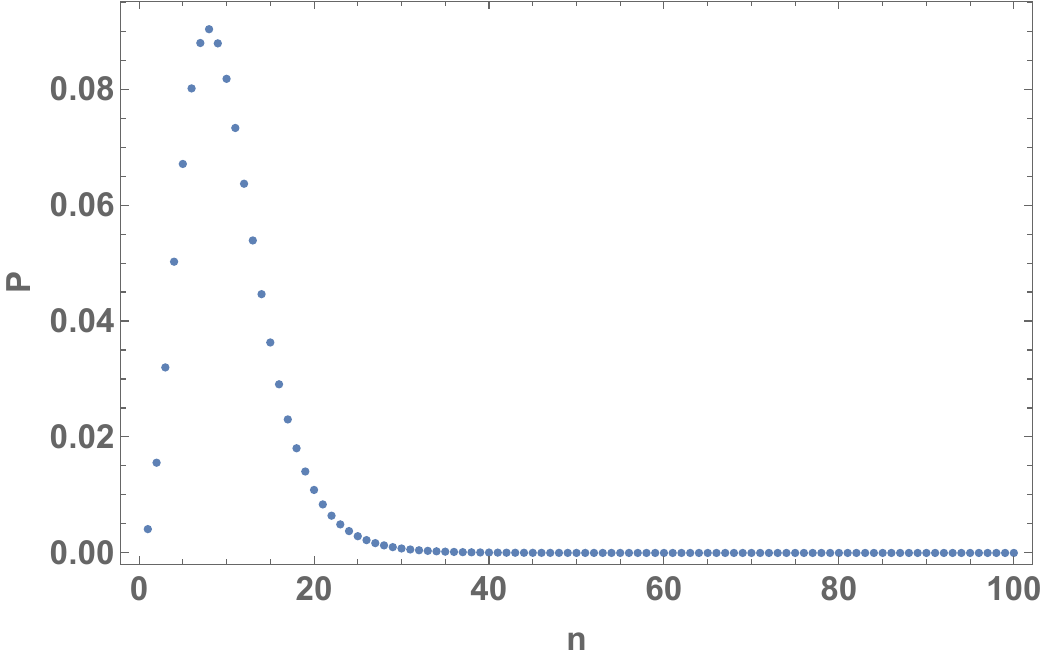}}
	\subcaptionbox{$m = 2 \times {10^{ - 17}}kg$\label{p2}}
	{\includegraphics[scale=0.4]{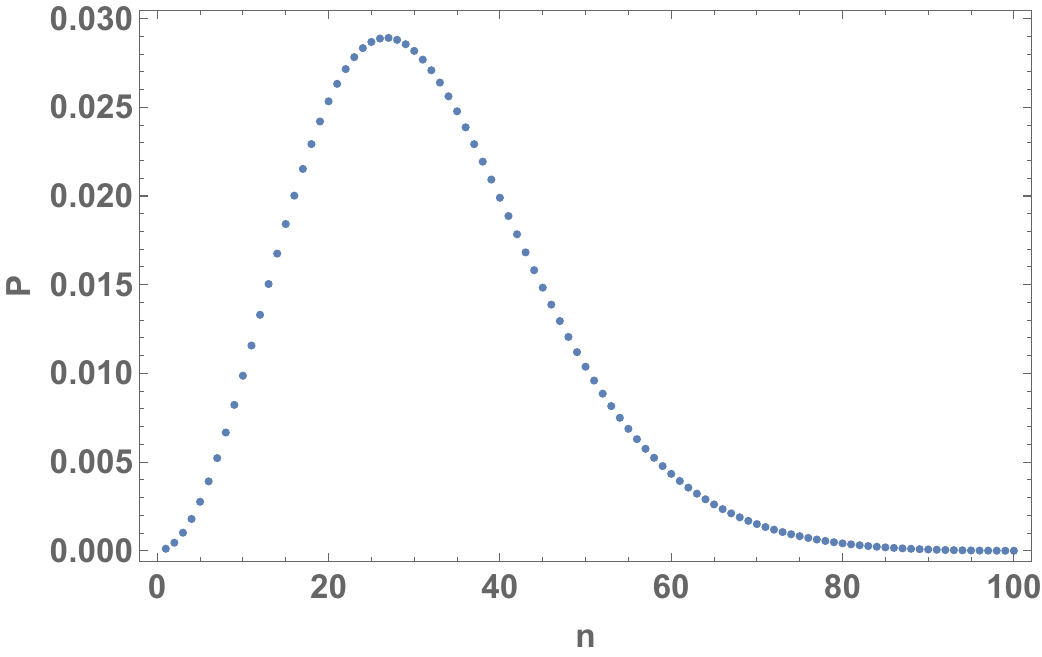}}
	\subcaptionbox{$m = 3 \times {10^{ - 17}}kg$\label{p3}}
	{\includegraphics[scale=0.4]{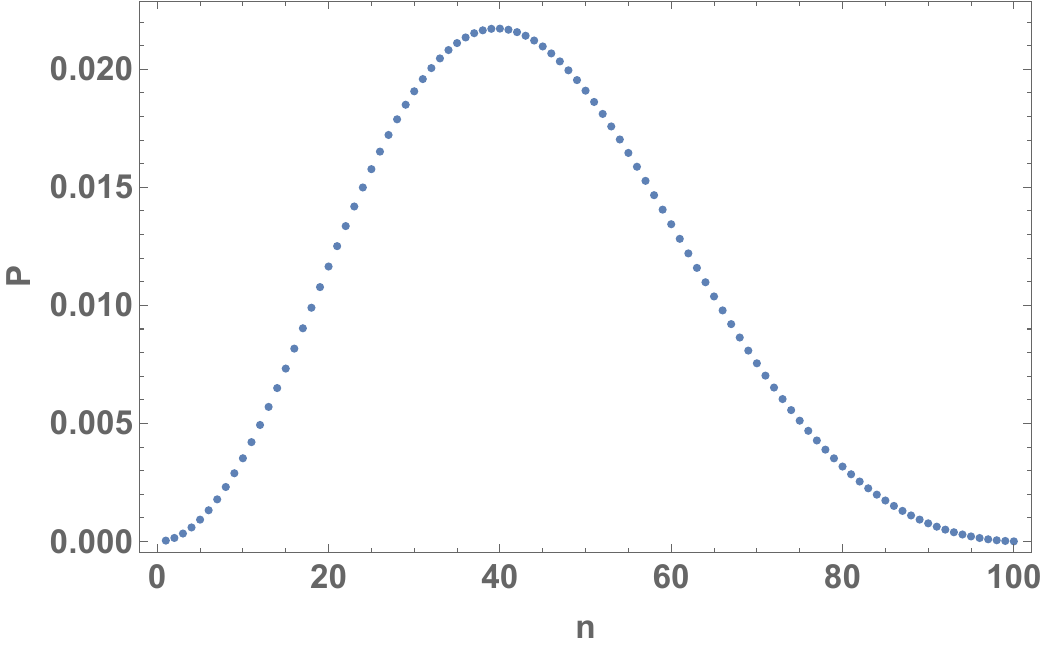}}
	\caption{ The probability distribution as a function of the particle mass, where $L$ is fixed to $50\mu$m, $d$ is fixed to $1\mu$m.}\label{p4}
\end{figure}

The probability distribution of the components of a given particle, such as A, in the joint eigenstate of particles A and B is calculated as follows:
\beq
{P_n} = {\rm{Tr}}\left( \hat \rho_A \left| n \right\rangle \left\langle n \right| \right)
\eeq
We take $L=50 \mu m$ and $d=1\mu$m for example, and the probability distribution varies with the mass of particles m as shown in Fig.\ref{p4}.

The probability distribution $P \approx 0$ around the high energy levels near $n=100$ implies that the components of these high energy levels are redundant. All parameters in this paper are ensured to maintain such a redundant relationship, ensuring that $n=100$ for our numerical calculations to be sufficiently accurate.	

  
  \end{appendix}

\end{document}